\documentclass[aps,prl,twocolumn,showpacs,preprintnumbers,superscriptaddress,amsmath,amssymb]{revtex4-1}

\usepackage{graphicx, times}
\usepackage{epstopdf}
\usepackage{bm}

\begin{document}

\title{Experimental Observations of Group Synchrony in a System of
  Chaotic Optoelectronic Oscillators}

\author{Caitlin R.
  S.
  Williams} \affiliation{Institute for Research in Electronics and
  Applied Physics, University of Maryland, College Park, MD 20742}
\affiliation{Department of Physics, University of Maryland, College
  Park, MD 20742}

\author{Thomas E.
  Murphy } \affiliation{Institute for Research in Electronics and
  Applied Physics, University of Maryland, College Park, MD 20742}
\affiliation{Department of Electrical and Computer Engineering,
  University of Maryland, College Park, MD 20742}

\author{Rajarshi Roy} \affiliation{Institute for Research in
  Electronics and Applied Physics, University of Maryland, College
  Park, MD 20742} \affiliation{Department of Physics, University of
  Maryland, College Park, MD 20742} \affiliation{Institute for
  Physical Science and Technology, University of Maryland, College
  Park, MD 20742}

\author{Francesco Sorrentino} \affiliation{Institute for Research in
  Electronics and Applied Physics, University of Maryland, College
  Park, MD 20742} \affiliation{Department of Mechanical Engineering,
  University of New Mexico, Albuquerque, New Mexico 87131, USA}

\author{Thomas Dahms} \affiliation{Institut f\"{u}r Theoretische
  Physik, Technische Universit\"{a}t Berlin, 10623 Berlin, Germany}

\author{Eckehard Sch\"{o}ll} \affiliation{Institut f\"{u}r
  Theoretische Physik, Technische Universit\"{a}t Berlin, 10623
  Berlin, Germany}

\date{\today}

\begin{abstract}
  We experimentally demonstrate group synchrony in a network of four
  nonlinear optoelectronic oscillators with time-delayed coupling.
  We divide the nodes into two groups of two each, by giving each
  group different parameters and by enabling only inter-group
  coupling.
  When coupled in this fashion, the two groups display different
  dynamics, with no isochronal synchrony between them, but
  the nodes in a single group are isochronally synchronized, even
  though there is no intra-group coupling.
  We compare experimental behavior with theoretical and numerical
  results.

\end{abstract}

\maketitle

The last years have seen a vast increase in the interest in coupled
dynamical systems, ranging from few coupled elements to complex networks
\cite{Albert2002RMP,BOC06a}.
Besides the focus on network structure and topology, the research area of
synchronization in networks has grown rapidly \cite{PEC90,PIK01}.
The groundbreaking work on the {\em master stability function} (MSF) by
Pecora and Carroll has bridged the gap between topology and dynamics
by allowing predictions about synchronization based solely on
the nodes' dynamics and the eigenvalue spectrum of the coupling matrix
\cite{PEC98}.

While the MSF theory was originally developed for identical, isochronous
synchronization, more complex patterns of synchronization are observed
in applications in, e.g., neural systems, genetic regulation, or optical systems
\cite{Buldu2007JoLT,Gonzalez2007EPL,Kestler2007PRE,Aviad2012OptExp,Nixon2011PRL,Amann2008JoP,
PIG07,JEN09,Choe2010PRE,Kanter2011EPL}.
These patterns include, for example, sublattice synchronization in coupled loops of identical oscillators with heterogeneous delays \cite{Kanter2011EPLb}, pairwise synchronization of pairwise identical nodes coupled through a common channel
\cite{Kestler2008PRE}, and more general group synchronization \cite{Sorrentino2007PRE}.
{\em In group synchronization the local dynamics in synchronized clusters
can be different from the dynamics in the other cluster(s)}, which extends the possibility of synchronization behavior to networks formed of heterogeneous dynamical systems, as they appear in a variety of
applications. Moreover, these synchronous patterns can be observed even when there is no intra-group coupling.
Sorrentino and Ott have generalized the MSF approach to group
synchronization \cite{Sorrentino2007PRE}, and recent work by Dahms {\em et al.}
considers time-delayed coupling of an arbitrary number of groups \cite{Dahms2012PRE}.

In this Letter, we demonstrate the successful realization of group
synchronization of chaotic dynamics in an array of four optoelectronic oscillators.
Optoelectronic oscillators with time-delayed feedback have been found
to show a multitude of different dynamical
behaviors ranging from steady-state to chaotic dynamics depending on
parameters
\cite{Chembo2005PRL,PEI09,Murphy2010PTRSA,Callan2010PRL,Rosin2011EPL,Ravoori2011PRL}. In this work we experimentally demonstrate group synchrony, where the two groups display different fluctuation amplitudes.  Remarkably, the two groups of synchronized oscillators are not directly coupled to each other; they are only coupled to those of the other groups.

\begin{figure} \centering
  \includegraphics[width=3.25in]{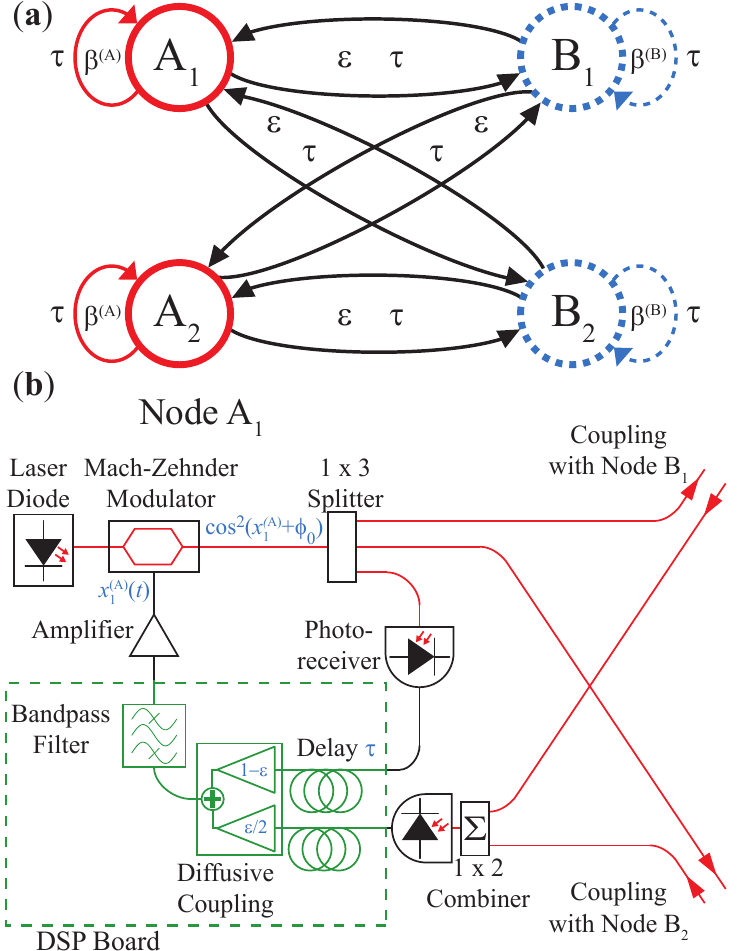}
  \caption{(Color online)  (a) Schematic of four nodes separated into two groups, A
    (red, solid) and B (blue, dashed).
    (b) Experimental setup of a single node, showing coupling to the
    other nodes according to the configuration in (a).
    } \label{NetworkSchematics}
\end{figure}

The experimental setup consists of four optoelectronic feedback loops,
which act as the four nodes of the network. We consider several coupling schemes. In the first one, the nodes are coupled
together in the configuration shown in Fig.~\ref{NetworkSchematics}(a)
in order to form two groups.  There are no direct coupling links
between two nodes in the same group.  However, a node is coupled
bidirectionally to both of the nodes in the other group.  In this
experiment, the coupling strength, $\varepsilon$, and coupling delay,
$\tau$, are the same for all coupling links. However, the parameters
of the nodes differ depending on which group the nodes are in.  Both
of the nodes in group A are identical, and both of the nodes in group
B are identical, but the nodes in group A are not identical to the
nodes in group B.  In Fig.~\ref{NetworkSchematics}(a), the coupling
links are shown in black (arrows in each direction to indicate
bidirectional coupling), and the self-feedback of the nodes is
indicated by the gray (colored) lines and arrows.

A schematic of a single node is shown in
Fig.~\ref{NetworkSchematics}(b), where red lines indicate optical
fibers, and black or green lines indicate electronic paths.  In each
node, light from a diode laser passes through a Mach-Zehnder modulator
(MZM), whose output light intensity is $\cos^2(x+\phi_0)$ for an input
voltage signal $x$.  There is a controllable bias phase of the MZM,
which we set to be $\phi_0 = \frac{\pi}{4}$.  The optical signal is
split into three equal signals: one is the feedback signal, and the
other two are the coupling to the two nodes in the opposite group.  A
photoreceiver converts the feedback optical signal to an electrical
signal, which is one of the two inputs to the DSP (digital signal
processing) board.  The incoming optical signals from the two nodes of
the other group are combined optically before a second photoreceiver
converts the composite coupled signal to an electronic signal, which
is the second input of the DSP board.  The DSP board implements the
feedback and coupling time delays, which are the same for this
experiment ($\tau = 1.4$ ms), and a diffusive coupling scheme. The
feedback signal is scaled by a factor of $1-\varepsilon$, while each incoming signal to a node is scaled by a factor of
$\frac{\varepsilon}{n_{\text{in}}}$, for the global coupling strength,
$\varepsilon$, and the number of links incoming to a node, $n_{\text{in}}$.  For the configuration shown in Fig.~\ref{NetworkSchematics}(a), $n_{\text{in}}=2$ for all nodes, but in general, $n_{\text{in}}$ can be different for each node. The feedback and coupled signals are summed on the DSP
board.

The DSP board also implements a digital filter, which is a two-pole
bandpass filter with cutoff frequencies at  100 Hz and 2.5 kHz and a sampling rate of 24
kSamples/s, and also scales the combined signal by a scale factor,
which controls the feedback strength, which we denote $\beta$.
The output of the DSP board is amplified with a voltage amplifier,
whose output drives the MZM.
Although $\beta$ is a combination of gains of the photoreceiver,
amplifier, and other components, the DSP board is the only place where
the gain is changed.

For this experiment, all parameters except for $\beta$ are identical
in all four nodes.
We keep $\beta$ identical among the members of each group but allow a
different $\beta$ for each group, denoted by $\beta^{(A)}$ and
$\beta^{(B)}$.
Previous studies have revealed the wide variety of behaviors that are
possible for this type of system, depending on the value of
$\beta$~\cite{Murphy2010PTRSA}.
For this study, we have used a range of $\beta$ from 0 to 10, with the
experiments focusing on cases of $\beta > 3$, for which the system
displays chaos (with some periodic windows) when the nodes are not coupled.

\begin{figure} \centering
  \includegraphics[width=3.375in]{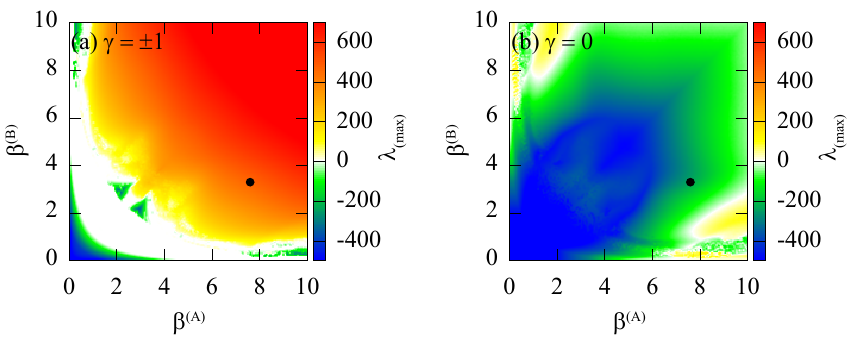}
  \caption{(Color online) Maximum Lyapunov exponent
    $\lambda_{\textnormal{max}}$ as a function of $\beta^{(A)}$ and
    $\beta^{(B)}$: (a) in the longitudinal directions $\gamma=\pm 1$,
    (b) in the transverse direction $\gamma=0$. White areas correspond
    to $\lambda_{\textnormal{max}}=0$.  Black dot indicates values of
    $\beta^{(A)}$ and $\beta^{(B)}$ used in this experiment.
  } \label{Stability}
\end{figure}

For each run of the experiment, the nodes are started from random
initial conditions.
This system has a time delay, so the initial condition will be a
function of time.
Thus, we record the random electrical activity at the input to the DSP
in the absence of coupling and feedback for 1 second to provide the
initial states for the nodes.
After recording an initial condition, we enable feedback for 4 seconds,
which is long enough for transients to disappear.
At the end of this period, we enable coupling.
Data are taken after transients have died out.

The system of coupled feedback loops can be well-described by a
mathematical model with a system of time delay differential
equations for the voltages input to the MZMs $x_i^{(m)} \in \mathbb{R}$ and the vectors describing the states of the filters
$\mathbf{u}_i^{(m)} \in \mathbb{R}^2$~\cite{Murphy2010PTRSA}:
\begin{equation}
  \mathbf{\dot{u}}_i^{(m)}(t) = \mathbf{E} \mathbf{u}_i^{(m)}(t)-\mathbf{F} \beta^{(m)} \cos^2(x_i^{(m)}(t-\tau)+\phi_0),
  \label{EoM1}
\end{equation}
\begin{equation}
  x_i^{(m)}(t)= \mathbf{G} [\mathbf{u}_i^{(m)}(t)+ \varepsilon\sum_j K_{ij}^{(m)} (\mathbf{u}_j^{(m')}(t)-\mathbf{u}_i^{(m)}(t))],
  \label{EoM2}
\end{equation}
where $m$ and $m'\not=m$ denote the groups A or B,
and $i$ indicates the node within a group.
$\mathbf{E}=(\begin{smallmatrix}
  -(\omega_H + \omega_L) & -\omega_L \\
  \omega_H & 0
\end{smallmatrix})$,
$\mathbf{F}= (\begin{smallmatrix}
  \omega_L \\
  0
\end{smallmatrix})$, and
$\mathbf{G}=(\begin{smallmatrix} 1 & 0
\end{smallmatrix})$
are constant matrices that describe the filter. The filter parameters
are chosen as $\omega_L=2\pi \cdot 2.5$~kHz and $\omega_H=2\pi \cdot
0.1$~kHz. For a bipartite network with no intra-group coupling, we define the inter-group coupling matrices
$\mathbf{K}^{(m)}=\{K^{(m)}_{ij}\}$:
\begin{equation}
\mathbf{K} = \left(
  \begin{array}{cc}
    0 & \mathbf{K}^{(A)}\\
    \mathbf{K}^{(B)} & 0\\
  \end{array}
  \right),
  \label{Kgeneral}
\end{equation}
where $\mathbf{K}$ is the overall coupling matrix for the entire network.
For the configuration shown in Fig.~\ref{NetworkSchematics}(a), $i,j=1,2$, and $\mathbf{K}^{(A)}=\mathbf{K}^{(B)} = \frac{1}{2}(
  \begin{smallmatrix}
    1 & 1\\
    1 & 1\\
  \end{smallmatrix})$
so that 
\begin{equation}
  \mathbf{K} = \left(
  \begin{array}{cc}
    0 & \mathbf{K}^{(A)}\\
    \mathbf{K}^{(B)} & 0\\
  \end{array}
  \right)=
   \frac{1}{2}\left(
  \begin{array}{cccc}
    0 & 0 & 1 & 1\\
    0 & 0 & 1 & 1\\
    1 & 1 & 0 & 0\\
    1 & 1 & 0 & 0\\
  \end{array}
\right).
\label{Kfull}
\end{equation}
Equations~\eqref{EoM1} and \eqref{EoM2} can describe the dynamics of the uncoupled nodes if we set the coupling strength $\varepsilon=0$, as
the second term in Eq.~\eqref{EoM2} represents the diffusive coupling
scheme.

Numerical simulations use a discrete-time implementation of these
differential equations, as described in Ref.~\cite{Murphy2010PTRSA}.
The simulations of uncoupled and coupled systems are in excellent
agreement with the experimental results for the variety of dynamical
behaviors that can be observed.

\begin{figure} \centering
  \includegraphics[width=3.375in]{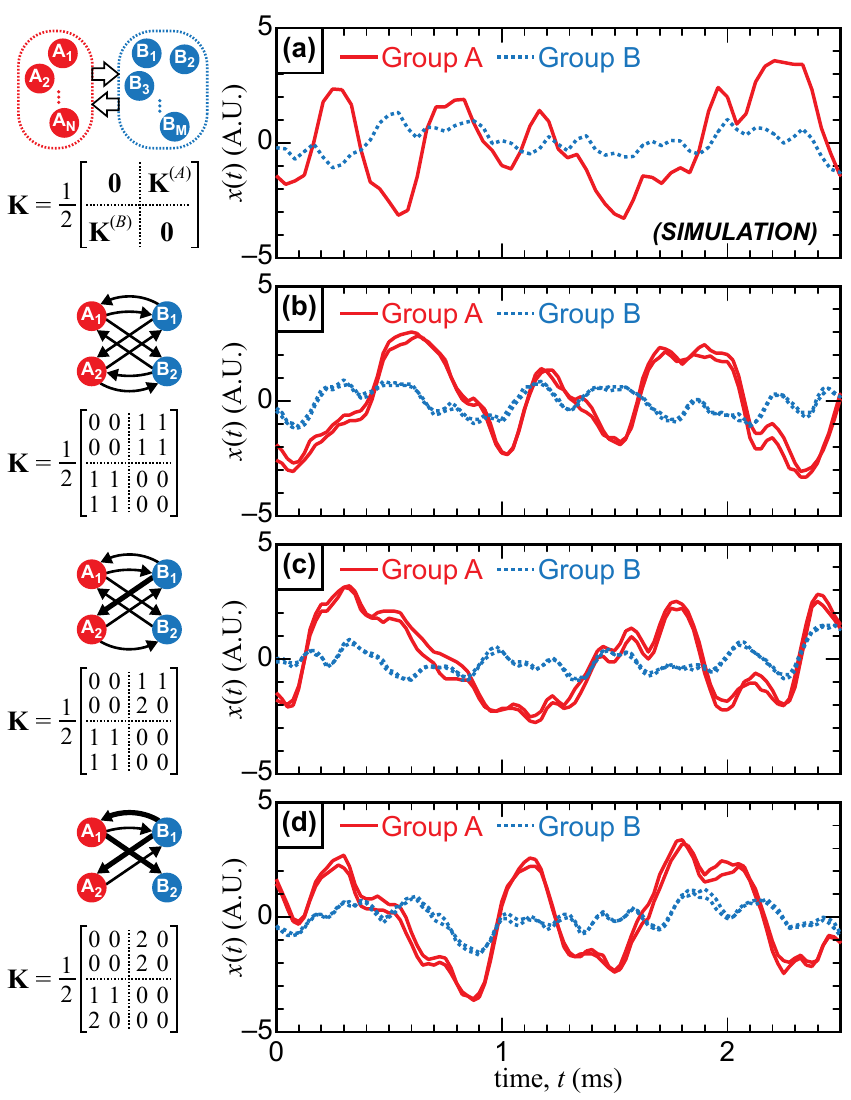}
  \caption{(Color online) (a) Simulated motion in the synchronization manifold, obtained by numerically solving Eqs.~\eqref{Synchronized1} and \eqref{Synchronized2}, showing the predicted group-synchronous state.  (b),(c),(d) Experimentally measured time traces from three different network configurations (indicated by the coupling scheme and the coupling matrices $\mathbf{K}$) that achieve group synchrony.  All three networks have the same eigenspectra, but the configuration in (b) is symmetric while those in (c) and (d) are not.
    } \label{TimeTrace}
\end{figure}

We will now investigate the existence and stability of the group synchronous
solution, i.e., we will derive analytical conditions determining whether such a solution (in which the two nodes of each group are identically and isochronously
synchronized, but there is no identical synchrony between nodes of different
groups) exists for given values of $\beta^{(A)}$ and $\beta^{(B)}$,
and if it does, if that solution is stable.
We use the approach described in~\cite{Sorrentino2007PRE,
  Dahms2012PRE}. The condition for the existence of the group synchronous solution for a particular coupling configuration is that
  \begin{equation}
  \sum_j K_{ij}^{(m)}=c^{(m)}, m=\{A,B\}, 
  \label{RowSum}
  \end{equation}
i.e., that the row sum of the matrices $\mathbf{K}^{(m)}$ is constant.  For the work reported here, we fix $c^{(A)}=c^{(B)}=1$.

\begin{figure} \centering
  \includegraphics[width=3.375in]{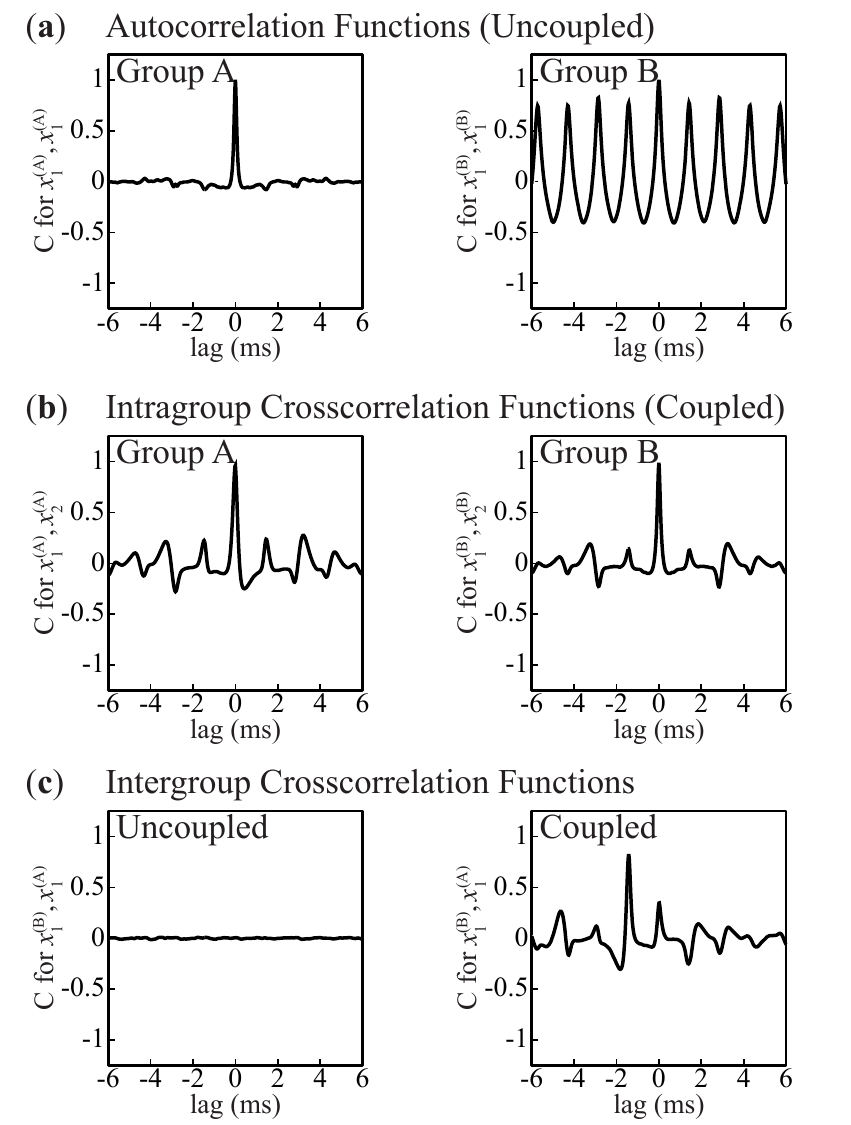}
  \caption{Correlation functions of 3.9 s of experimental data.  (a) Autocorrelation functions for the dynamics of group A (left) and group B (right), with no coupling in the system.  (b) Cross-correlation functions between the two nodes in group A (left) and group B (right), for coupled nodes. (c) Cross-correlation functions between one node in group A and one node in group B for the uncoupled system (left) and coupled system (right).}
  \label{AttractorsCorrelations}
\end{figure}

The group synchronized dynamics for group~\textit{m} is given by
\begin{equation}
  \label{Synchronized1}
  \dot{\mathbf{u}}_s^{(m)}(t) = \mathbf{E} \mathbf{u}_s^{(m)}(t)\\-\mathbf{F} \beta^{(m)} \cos^2(x_s^{(m)}(t-\tau)+\phi_0),
\end{equation}
\begin{equation}
  x_s^{(m)}(t)= \mathbf{G} [(\mathbf{u}_s^{(m)}(t)+ \varepsilon (\mathbf{u}_s^{(m')}(t)-\mathbf{u}_s^{(m)}(t))].
  \label{Synchronized2}
\end{equation}
Linearizing Eqs.~\eqref{EoM1} and \eqref{EoM2} about
the synchronous solution $\mathbf{u}_s^{(m)}$ ($m=A,B$), we obtain
the master stability equations:
\begin{equation}
  \label{MS}
\begin{split}
  \delta \dot{\mathbf{u}}^{(m)}(t) =& \mathbf{E}\delta \mathbf{u}^{(m)}(t)-\mathbf{F}\beta^{(m)}\sin(2x_s^{(m)}(t-\tau)+2\phi_0)\\
  &\times\mathbf{G}[(1-\varepsilon)\delta \mathbf{u}^{(m)}(t-\tau)+\varepsilon\gamma \mathbf{u}^{(m')}(t-\tau)].
\end{split}
\end{equation}
In Eq.
\eqref{MS}, $\gamma$ is a parameter that is chosen from the eigenvalue
spectrum of $\mathbf{K}$. The largest Lyapunov exponent
as a function of this parameter $\gamma$ is called the MSF.   For the configurations presented here, the nonzero eigenvalues of $\mathbf{K}$ are 1 and -1, and any remaining eigenvalues are  zeros.  Therefore, the stability results will be identical for any two-group network whose nodes are described by Eqs.~\eqref{EoM1} and \eqref{EoM2} and whose coupling matrix is in the form of \eqref{Kgeneral}, satisfies \eqref{RowSum}, and has identical rows for either $\mathbf{K}^{(A)}$ or $\mathbf{K}^{(A)}$ (for a proof, see supplemental material).

The eigenvalues $\gamma=-1$ and $\gamma=1$ in the master stability equation~\eqref{MS} correspond to
perturbations parallel to the synchronization manifold.  The
corresponding value of the MSF determines the dynamical behavior
inside the synchronization manifold and is shown in
Fig.~\ref{Stability}(a) in dependence on the parameters $\beta^{(A)}$
and $\beta^{(B)}$.  Negative, zero, and positive values denote
fixed-point, periodic, and chaotic dynamics, respectively.  Due to the
inversion symmetry of the MSF for two-group synchronization
~\cite{Sorrentino2007PRE,Dahms2012PRE}, the MSF values are identical
for $\gamma=-1$ and $\gamma=1$.

Transverse stability of the synchronization manifold is determined by
using the eigenvalue $\gamma=0$ in Eq.~\eqref{MS}.
Figure~\ref{Stability}(b) shows the largest Lyapunov exponent in the
transverse direction, which is negative for almost the entire range of
$\beta^{(A)}$ and $\beta^{(B)}$ that is shown, indicating that we
expect the group synchronous solution to be stable for most
parameters.

To observe group synchrony in this system, we select dissimilar values
of $\beta^{(A)}$ and $\beta^{(B)}$, as shown by the black dots in Fig.~\ref{Stability}.
The global coupling strength is chosen as $\varepsilon=0.8$.
The experimental values for $\beta^{(A)}$ and $\beta^{(B)}$ were
adjusted using the DSP board.
The values of $\beta^{(A)}$ and $\beta^{(B)}$ used in simulation were
established by varying the values close to the experimental values to
find nearby values which match best the dynamical behavior of the experiments
for uncoupled nodes, obtained from the shape of the reconstructed attractor in phase space.
Since the values determined experimentally as $\beta^{(A)}=7.6$ and $\beta^{(B)} =3.3$ are subject to measurement uncertainties, it is not surprising that we find slightly different values in simulation, i.e., $\beta^{(A)} = 7.66$ and $\beta^{(B)} = 3.28$. Comparison of uncoupled and coupled time traces in experiment and simulation is shown in the supplemental material, Fig. \ref{TimeTraceSM}.

Figure~\ref{TimeTrace} shows experimental and simulated time traces of the coupled system.  The simulated traces in Fig.~\ref{TimeTrace}(a) show the behavior of \textit{any} two-group system displaying stable group synchrony according to Eqs.~\eqref{Synchronized1} and \eqref{Synchronized2}, with the parameters we have used here.  Figure~\ref{TimeTrace}(b) shows experimental results for a system coupled according to Fig.~\ref{NetworkSchematics}(a).  These time traces show that there is identical, isochronal synchrony between $x_1^{(A)}$ and $x_2^{(A)}$, and between $x_1^{(B)}$ and $x_2^{(B)}$, but not identical synchrony between the groups.  Thus, this is an example of group synchrony.  We also performed experiments on two asymmetric four-node configurations.  These configurations were created by removing links from the original structure of Fig.~\ref{NetworkSchematics}(a), while preserving the constant row sum and eigenvalues (1, -1, 0, and 0) of $\mathbf{K}$, keeping all other parameters the same.  Their topologies and dynamics are shown in Figs.~\ref{TimeTrace}(c) and \ref{TimeTrace}(d).  Because these schemes are also described by Fig.~\ref{Stability}, they also display group synchrony.  In the experimental time traces, there are slight differences between the two traces of one group, due to the intrinsic experimental noise and mismatch we expect in any real system.  An example of a larger network that displays the same behavior is presented in the supplemental material, Fig. \ref{SevenNodes}.

To further examine the nature of the synchrony of this system, we calculate the correlation functions of the experimental time traces, as shown in Fig.~\ref{AttractorsCorrelations} for the topology shown in Fig.~\ref{NetworkSchematics}(a).  For two variables $y(t)$ and $z(t)$, which each have a mean of zero, we define the correlation function $C$ as a function of time lag $\Delta t$ \cite{MUL04}:
\begin{equation}
\label{Correlation}
 C(\Delta t) = \frac{\langle y(t) z(t+\Delta t) \rangle}{\sqrt {\langle y^2(t) \rangle \langle z^2(t) \rangle}}.
 \end{equation}
 Figure~\ref{AttractorsCorrelations}(a) shows the autocorrelation
 functions for one node in each group when the nodes are uncoupled.
 The autocorrelation of $x_1^{(A)}$ shows only a peak at zero time
 lag, which indicates chaotic dynamics, while the autocorrelation of
 $x_1^{(B)}$ shows periodic dynamics, with correlation peaks at
 intervals of the time delay $\tau=1.4$ ms.  In
 Fig.~\ref{AttractorsCorrelations}(b), we show the cross-correlation
 functions of $x_1^{(A)}$ with $x_2^{(A)}$, and of $x_1^{(B)}$ with
 $x_2^{(B)}$ for the coupled system, which confirms identical,
 isochronal chaotic synchronization between the two nodes in a single
 group.  Figure ~\ref{AttractorsCorrelations}(c) shows the
 cross-correlation functions between two nodes in different groups,
 without and with coupling.  The uncoupled case has no correlation, as
 we expect, but the coupled case has a high correlation peak at a lag
 of $\Delta t=-1.4$ ms.  From this, we can see that there is
 time-lagged phase synchrony between the two groups, with the dynamics
 of group B leading the dynamics of group A by the system delay,
 $\tau$.  However, the amplitudes of fluctuations of the two groups
 are still different after coupling, so there is no complete
 synchronization, and we have an interesting situation of the
 simultaneous coexistence of intragroup isochronal identical synchrony
 and time-lagged phase synchrony between the groups.

In conclusion, we have examined a four-node system of nonlinear
optoelectronic oscillators in the case where there are two groups of
nodes with dissimilar parameters.  Our experiments display the
phenomenon of group synchronization, and we analyze the stability of
the group synchronized solutions for chaotic dynamical states. It is
remarkable that, although the coupling is entirely between the
different groups and not within the groups, identical isochronal
synchronization within each group is induced by this coupling, while
the two groups are not mutually amplitude synchronized, as predicted
by our stability analysis using the generalized master stability
function \cite{Sorrentino2007PRE,Dahms2012PRE}. Thus the nodes of
group B act as a kind of dynamical relay \cite{FIS06} for the nodes of
group A, and vice versa.  These results have been experimentally demonstrated with three coupling configurations, and conditions for observing group synchrony in other networks have been discussed.

Our observations go beyond previous work on sublattice and cluster synchrony, where the experiments focused
 on optical phase synchronization for coupled lasers without
self-feedback \cite{Aviad2012OptExp,Nixon2011PRL}. Group
synchronization in larger networks is a significant challenge for
future experimental investigation.

This work was supported by DOD MURI grant ONR N000140710734 and by DFG
in the framework of SFB 910.  The authors would like to acknowledge helpful comments from I. Kanter and L. Pecora.

\setcounter{figure}{0}
\setcounter{equation}{0}

\onecolumngrid
\clearpage

\section{Supplemental Material}
Figures \ref{TimeTrace}(b), \ref{TimeTrace}(c), and \ref{TimeTrace}(d) show that stable group synchrony is experimentally observed for three different coupling configurations. Here we show that our stability analysis and the numerical computations in Fig. \ref{Stability} apply to all of these coupling schemes and, more generally, to a whole class of networks, characterized by an arbitrary number of nodes in both the groups $A$ and $B$.

We define $N_A$ and $N_B$ the number of nodes in group $A$ and $B$, respectively. Then the couplings are fully described by the $N_A \times N_B$ coupling matrix $\mathbf{K}^{(A)}$, whose entries $\{ K^{(A)}_{ij} \}$ represent the intensity of the direct interaction
from system  $j$ in group $B$ to  $i$ in group $A$ and  the $N_B \times N_A$ matrix $\mathbf{K}^{(B)}$, whose entries $\{ K^{(B)}_{ij} \}$ represent the intensity of the direct interaction
from system  $i$ in group $A$ to $j$ in group $B$.

First we note that the motion in the synchronization manifold (Eqs. \eqref{Synchronized1} and \eqref{Synchronized2}) applies to any network described by Eqs. \eqref{EoM1} and \eqref{EoM2}, as long as the entries along the rows of the matrices $\mathbf{K}^{(A)}$ and $\mathbf{K}^{(B)}$ sum to one. Thus in what follows, we will limit our attention to the case that
\renewcommand{\theequation}{{S\arabic{equation}}}
\begin{equation} \label{cacb}
c^{(A)}=c^{(B)}=1.
\end{equation} 
If assumption (\ref{cacb}) is verified, it follows that the maximum Lyapunov exponent of the synchronous solution shown in Fig. \ref{Stability}(a) does not depend on the details of the underlying network structure.

According to Ref. \cite{Sorrentino2007PRE}, a master stability function approach to group synchronization is possible for any network described by Eqs. \eqref{EoM1} and \eqref{EoM2}, under the assumption (\ref{cacb}). For any such network stability depends on the eigenvalues of the matrix
\begin{equation}
\mathbf{K}=\left(
  \begin{array}{cc}
    0 & \mathbf{K}^{(A)} \\
    \mathbf{K}^{(B)} & 0 \\
  \end{array}
\right).
\end{equation}

From Ref. \cite{Sorrentino2007PRE} we see that these eigenvalues are
\begin{equation}
\Lambda= [0,0,...,0] \bigcup [\pm \sqrt{\tilde{\lambda}_1}, \pm \sqrt{\tilde{\lambda}_2},...,\pm \sqrt{\tilde \lambda_{{N}_{min}}}], \label{quattordici}
\end{equation}
where $[0,0,...,0]$ denotes $|N_A-N_B|$ zeros and $\tilde \Lambda=\{ \tilde {\lambda}_1,..., \tilde \lambda_{{N}_{min}} \}$ denotes the spectrum of the $N_{min} \times N_{min}$ matrix,
\begin{align}
D= \left\{ \begin{array} {ccc} {\mathbf{K}^{(A)} \mathbf{K}^{(B)},} \quad \mbox{if} \quad {N_A \leq N_B,} \\ {\mathbf{K}^{(B)} \mathbf{K}^{(A)},}  \quad \mbox{if} \quad    {N_B < N_A,} \end{array} \right.
\end{align}

We proceed now under the assumption that either one of the two following conditions is satisfied,
\begin{subequations} \label{ab}
\begin{align}
\mathbf{K}^{(A)}=\left(
  \begin{array}{cccc}
    a \\
    a \\
    ... \\
    a
  \end{array}
\right) \label{ab_a},\\
\mathbf{K}^{(B)}=\left(
  \begin{array}{cccc}
    b \\
    b \\
    ... \\
    b
  \end{array}
\right)\label{ab_b},
\end{align}
\end{subequations}
where $a$ ($b$) is any $N_B$-dimensional row-vector ($N_A$-dimensional row-vector) with its entries summing to one. Then the matrix $D$ is in the form
\begin{equation}
D=\left(
  \begin{array}{cccc}
    d \\
    d \\
    ... \\
    d
  \end{array}
\right),
\end{equation}
where $d$ is an $N_{min}$-dimensional row-vector with sum of its entries equal one. Note that the underlying assumption is that either one of the matrices $\mathbf{K}^{(A)}$ and $\mathbf{K}^{(B)}$ is in the  form of Eq.\ (\ref{ab}) (not necessarily both). Then the eigenvalue equation for the matrix $D$ reduces to the following equation
\begin{equation} \label{eigd}
\lambda v_i = d \cdot v= \ell \quad \quad i=1,...,N_{min},
\end{equation}
where $v=[v_1,v_2,...,v_{N_{min}}]$ is an eigenvector, $\lambda$ is the associated eigenvalue,
and $\ell$ is a constant that does not depend on $i$.
There are two possible solutions for Eq.\ (\ref{eigd}):

(i) $\ell=0$. Then, $\lambda=0$ and $v$ is any vector orthogonal to $d$. There are $(N_{min}-1)$ such vectors.

(ii) $\ell \neq 0$. Then, $v_1=v_2=...=v_{N_{min}}$ and from the sum of the entries of the vector $d$ being equal one, it follows that $\lambda=1$.

Hence by using Eq. (\ref{quattordici}), and assuming satisfaction of either one of the Eqs. \eqref{ab_a} or \eqref{ab_b}, the eigenvalues of the matrix $K$ are
\begin{equation}
\Lambda= [0,0,...,0] \bigcup [-1,1],
\end{equation}
where here $[0,0,...,0]$ denotes $|N_A+N_B-2|$ zeros. It follows that stability of the group synchronous solution for any network that satisfies either Eqs. (\ref{ab}(a)) or (\ref{ab}(b))
is described by the plot in Fig. \ref{Stability}(b).

\newpage

\renewcommand{\thefigure}{{S\arabic{figure}}}
\begin{figure}[h]
\centering
\includegraphics{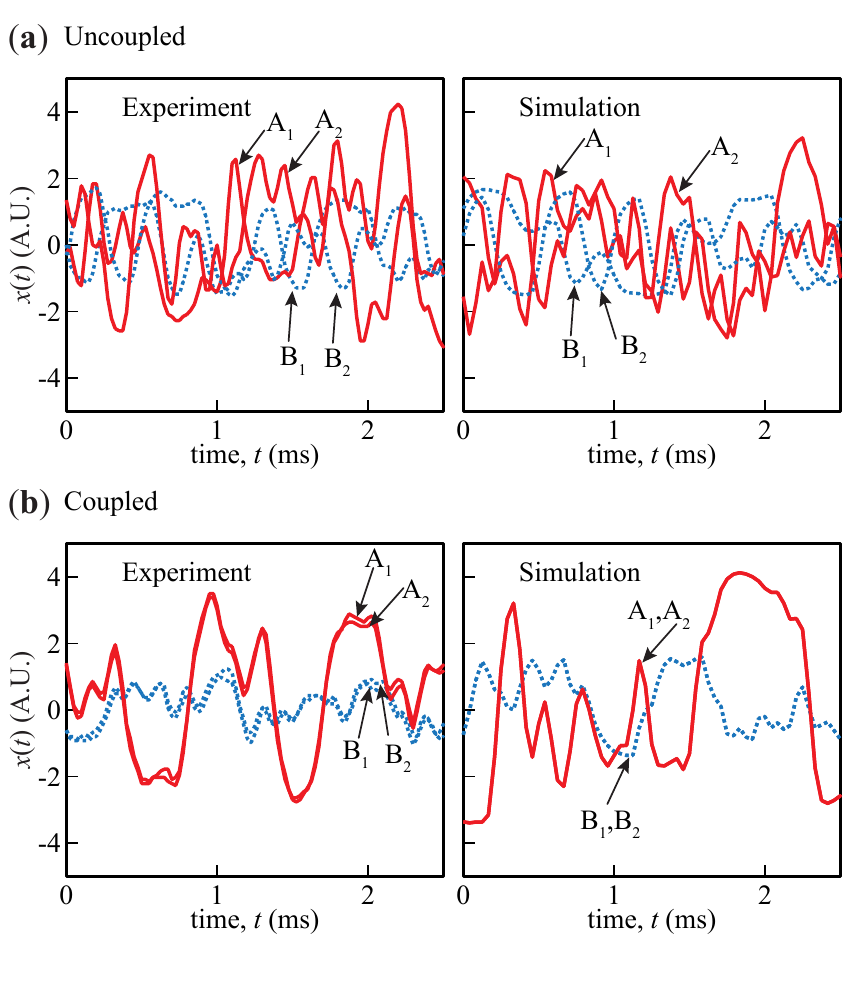}
\caption{Comparison of time traces from experiment (left column) and simulation
    (right column).
    Nodes in group A (B) are indicated by the red and solid (blue and
    dashed) lines.
    (a) Nodes are uncoupled, and the uncoupled nodes are completely unsynchronized.
    (b) Nodes are coupled according to the configuration in
    Fig.~\ref{NetworkSchematics}(a).  There is identical, isochronal synchrony between $x_1^{(A)}$ and $x_2^{(A)}$, and between $x_1^{(B)}$ and $x_2^{(B)}$, but not identical synchrony between the groups, so this is an example of group synchrony.  In the simulation, the two traces in one group are exactly synchronized and are indistinguishable, as we expect from a simulation without noise or mismatch.  In experiment, slight differences between synchronized traces arise from experimental noise and mismatch in the real experimental system.  In both the experiment and simulation, the dynamics of the nodes in group B have a significantly smaller amplitude than those in group A, with qualitatively similar dynamics between simulation and experiment.
    } \label{TimeTraceSM}
\end{figure}

\begin{figure}[h]
\centering
\includegraphics[width=6in]{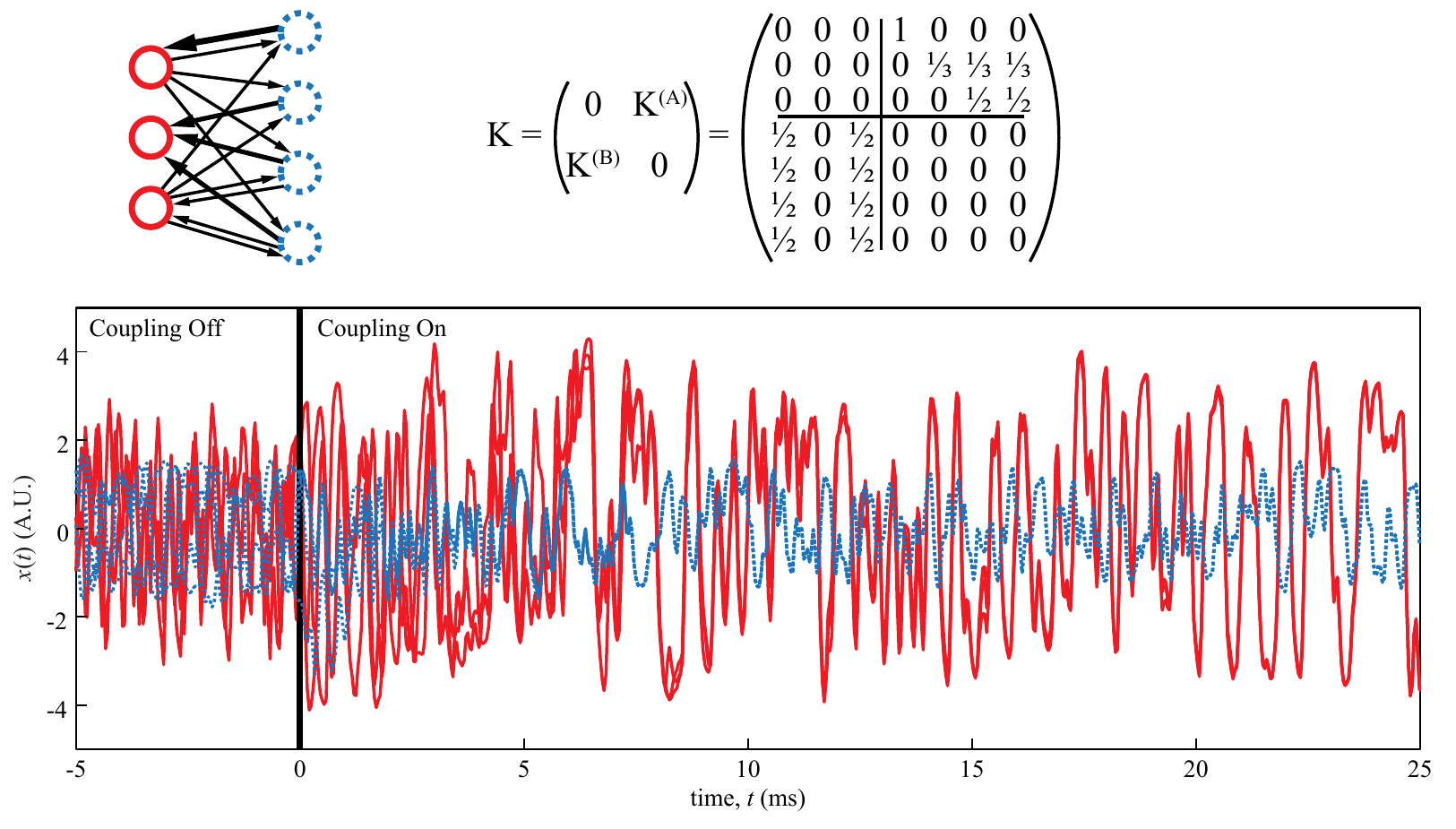}
\caption{Simulation of a seven node network displaying group synchrony.  The coupling scheme and  the associated coupling matrix are shown above the simulated time traces.  The parameters are the same as those presented in the letter.  Coupling is enabled at time $t=0$, and the nodes become group synchronized so that all three red, solid lines of Group A are identically synchronized, and all four blue, dashed traces of Group B are identically synchronized.  Note that the coupling matrix for the specific network configuration shown here satisfies Eq.~\eqref{ab_b}. Hence, the coupling matrix has eigenvalues -1, 1, and degenerate zeros. The stability of the group synchronous solution of this system is described by Fig.~\ref{Stability}(b).  The dynamics of the seven node, group synchronized system, are predicted by Fig.~\ref{Stability}(a), as illustrated by Fig.~\ref{TimeTrace}(a).}
\label{SevenNodes}
\end{figure}


\end{document}